# Least angle and $\ell_1$ penalized regression: A review[*][†]

## Tim Hesterberg, Nam Hee Choi, Lukas Meier, and Chris Fraley[§]

*Insightful Corp.[‡], University of Michigan, ETH Zürich, Insightful Corp.*

**Abstract:** Least Angle Regression is a promising technique for variable selection applications, offering a nice alternative to stepwise regression. It provides an explanation for the similar behavior of LASSO ($\ell_1$-penalized regression) and forward stagewise regression, and provides a fast implementation of both. The idea has caught on rapidly, and sparked a great deal of research interest. In this paper, we give an overview of Least Angle Regression and the current state of related research.

## Contents

[*]This work was supported by NIH SBIR Phase I 1R43GM074313-01 and Phase II 2R44GM074313-02 awards.
[†]This paper was accepted by Grace Wahba, Associate Editor for the IMS.
[‡]now at Google, Inc.
[§]corresponding author (fraley@insightful.com)

## 1. Introduction

> "I've got all these variables, but I don't know which ones to use."

Classification and regression problems with large numbers of candidate predictor variables occur in a wide variety of scientific fields, increasingly so with improvements in data collection technologies. For example, in microarray analysis, the number of predictors (genes) to be analyzed typically far exceeds the number of observations.

Goals in model selection include:

- accurate predictions,
- interpretable models—determining which predictors are meaningful,
- stability—small changes in the data should not result in large changes in either the subset of predictors used, the associated coefficients, or the predictions, and
- avoiding bias in hypothesis tests during or after variable selection.

Older methods, such as stepwise regression, all-subsets regression and ridge regression, fall short in one or more of these criteria. Modern procedures such as boosting (Freund and Schapire, 1997) forward stagewise regression (Hastie *et al.*, 2001), and LASSO (Tibshirani, 1996), improve stability and predictions.

Efron *et al.* (2004) show that there are strong connections between these modern methods and a method they call *least angle regression*, and develop an algorithmic framework that includes all of these methods and provides a fast implementation, for which they use the term 'LARS'. LARS is potentially revolutionary, offering interpretable models, stability, accurate predictions, graphical output that shows the key tradeoff in model complexity, and a simple data-based rule for determining the optimal level of complexity that nearly avoids the bias in hypothesis tests.

This idea has caught on rapidly in the academic community—a 'Google Scholar' search in May 2008 shows over 400 citations of Efron *et al.* (2004), and over 1000 citations of Tibshirani (1996).

We explain the importance of LARS in this introduction and in Section 2.1 and compare it to older variable selection or penalized regression methods in Section 2.2. We describe extensions in Section 3, alternate approaches in Section 4, and list some available software in Section 5.

## 2. History

### *2.1. Significance*

In 1996 one of us (Hesterberg) asked Brad Efron for the most important problems in statistics, fully expecting the answer to involve the bootstrap, given Efron's status as inventor. Instead, Efron named a single problem, *variable selection in regression*. This entails selecting variables from among a set of candidate variables, estimating parameters for those variables, and inference—hypotheses tests, standard errors, and confidence intervals.

It is hard to argue with this assessment. Regression, the problem of estimating a relationship between a response variable and various predictors (explanatory variables, covariates) is of paramount importance in statistics (particularly when we include "classification" problems, where the response variable is categorical). A large fraction of regression problems require some sort of choice of predictors. Efron's work has long been strongly grounded in solving real problems, many of them from biomedical consulting. His answer reflects the importance of variable selection in practice.

Classical tools for analyzing regression results, such as $t$ statistics for judging the significance of individual predictors, are based on the assumption that the set of predictors is fixed in advance. When instead the set is chosen adaptively, incorporating those variables that give the best fit for a particular set of data, the classical tools are biased. For example, if there are 10 candidate predictors, and we select the single one that gives the best fit, there is about a 40% chance that that variable will be judged significant at the 5% level, when in fact all predictors are independent of the response and each other. Similar bias holds for the $F$ test for comparing two models; it is based on the assumption that the two models are fixed in advance, rather than chosen adaptively.

This bias affects the variable selection process itself. Formal selection procedures such as stepwise regression and all-subsets regression are ultimately based on statistics related to the $F$ statistics for comparing models. Informal selection procedures, in which an analyst picks variables that give a good fit, are similarly affected.

In the preface to the second edition of *Subset Selection in Regression* (Miller 2002), Allan Miller noted that little progress had been made in the previous decade:

> What has happened in this field since the first edition was published in 1990?
>
> The short answer is that there has been very little progress. The increase in the speed of computers has been used to apply subset selection to an increasing range of models, linear, nonlinear, generalized linear models, to regression methods which are more robust against outliers than least squares, but we still know very little about the properties of the parameters of the best-fitting models chosen by these methods. From time-to-time simulation studies have been published, e.g. Adams (1990), Hurvich and Tsai (1990), and Roecker (1991), which have shown, for instance, that prediction errors using ordinary least squares are far too small, or that nominal 95% confidence regions only include the true parameter values in perhaps 50% of cases.

Problems arise not only in selecting variables, but also in estimating coefficients for those variables, and producing predictions. The coefficients and predictions are biased as well as unstable (small changes in the data may result in large changes in the set of variables included in a model and in the corresponding coefficients and predictions). Miller (2002) notes:

> As far as estimation of regression coefficients is concerned, there has been essentially no progress.

Least angle regression Efron *et al.* (2004), and its LASSO and forward stagewise variations, offer strong promise for producing interpretable models, accurate predictions, and approximately unbiased inferences.

## *2.2. LARS and Earlier Methods*

In this section we discuss various methods for regression with many variables, leading up to the original LARS paper (Efron *et al.*, 2004). We begin with "pure variable selection" methods such as stepwise regression and all-subsets regression that pick predictors, then estimate coefficients for those variables using standard criteria such as least-squares or maximum likelihood. In other words, these methods focus on variable selection, and do nothing special about estimating coefficients. We then move on to ridge regression, which does the converse—it is not concerned with variable selection (it uses all candidate predictors), and instead modifies how coefficients are estimated. We then discuss LASSO, a variation of ridge regression that modifies coefficient estimation so as to reduce some coefficients to zero, effectively performing variable selection. From there we move to forward stagewise regression, an incremental version of stepwise regression that gives results very similar to LASSO. Finally we turn to least angle regression, which connects all the methods.

We write LAR for least angle regression, and LARS to include LAR as well as LASSO or forward stagewise as implemented by least-angle methods. We use the terms predictors, covariates, and variables interchangeably (except we use the latter only when it is clear we are discussing predictors rather than response variables).

The example in this section involves linear regression, but most of the text applies as well to logistic, survival, and other nonlinear regressions in which the predictors are combined linearly. We note where there are differences between linear regression and the nonlinear cases.

TABLE 1

*Diabetes Study: 442 patients were measured on 10 baseline variables; a prediction model is desired for the response variable $Y$, a measure of disease progression one year after baseline. Predictors include age, sex, body mass index, average blood pressure, and six different blood serum measurements. One goal is to create a model that predicts the response from the predictors; a second is to find a smaller subset of predictors that fits well, suggesting that those variables are important factors in disease progression.*

| Patient | Age | Sex | BMI | BP | S1 | S2 | S3 | S4 | S5 | S6 | Y |
|--------:|----:|----:|----:|----:|----:|------:|----:|----:|----:|----:|----:|
| 1 | 59 | 2 | 32.1 | 101 | 157 | 93.2 | 38 | 4.0 | 4.9 | 87 | 151 |
| 2 | 48 | 1 | 21.6 | 87 | 183 | 103.2 | 70 | 3.0 | 3.9 | 69 | 75 |
| 3 | 72 | 2 | 30.5 | 93 | 156 | 93.6 | 41 | 4.0 | 4.7 | 85 | 141 |
| ⋮ | ⋮ | ⋮ | ⋮ | ⋮ | ⋮ | ⋮ | ⋮ | ⋮ | ⋮ | ⋮ | ⋮ |
| 442 | 36 | 1 | 19.6 | 71 | 250 | 133.2 | 97 | 3.0 | 4.6 | 92 | 57 |

### 2.2.1. Stepwise and All-Subsets Regression

We begin our description of various regression methods with stepwise and all-subsets regression, which focus on selecting variables for a model, rather than on how coefficients are estimated once variables are selected.

Forward stepwise regression begins by selecting the single predictor variable that produces the best fit, e.g. the smallest residual sum of squares. Another predictor is then added that produces the best fit in combination with the first, followed by a third that produces the best fit in combination with the first two, and so on. This process continues until some stopping criteria is reached, based e.g. on the number of predictors and lack of improvement in fit. For the diabetes data shown in Table 1, single best predictor is BMI; subsequent variables selected are S5, BP, S1, Sex, S2, S4, and S6.

The process is unstable, in that relatively small changes in the data might cause one variable to be selected instead of another, after which subsequent choices may be completely different.

Variations include backward stepwise regression, which starts with a larger model and sequentially removes variables that contribute least to the fit, and Efroymson's procedure (Efroymson, 1960), which combines forward and backward steps.

These algorithms are greedy, making the best change at each step, regardless of future effects. In contrast, all-subsets regression is exhaustive, considering all subsets of variables of each size, limited by a maximum number of best subsets (Furnival and Wilson, Jr., 1974). The advantage over stepwise procedures is that the best set of two predictors need not include the predictor that was best in isolation. The disadvantage is that biases in inference are even greater, because it considers a much greater number of possible models.

In the case of linear regression, computations for these stepwise and all-subsets procedures can be accomplished using a single pass through the data.

This improves speed substantially in the usual case in where are many more observations than predictors. Consider the model

$$\mathbf{Y} = \mathbf{X}\boldsymbol{\beta} + \boldsymbol{\epsilon} \tag{1}$$

where $\mathbf{Y}$ is a vector of length $n$, $\mathbf{X}$ an $n$ by $p$ matrix, $\boldsymbol{\beta}$ a vector of length $p$ containing regression coefficients, and $\boldsymbol{\epsilon}$ assumed to be a vector of independent normal noise terms. In variable selection, when some predictors are not included in a model, the corresponding terms in $\boldsymbol{\beta}$ are set to zero. There are a number of ways to compute regression coefficients and error sums of squares in both stepwise and all subsets regression. One possibility is to use the cross-product matrices $\mathbf{X}'\mathbf{X}$, $\mathbf{X}'\mathbf{Y}$, and $\mathbf{Y}'\mathbf{Y}$. Another is to use the $QR$ decomposition. Both the cross-product and $QR$ implementations can be computed in a single pass through the data, and in both cases there are efficient updating algorithms for adding or deleting variables. However, the $QR$ approach has better numerical properties. See e.g. Thisted (1988); Monahan (2001); Miller (2002) for further information.

For nonlinear regressions, the computations are iterative, and it is not possible to fit all models in a single pass through the data.

Those points carry over to LARS. The original LARS algorithm computes $\mathbf{X}'\mathbf{X}$ and $\mathbf{X}'\mathbf{Y}$ in one pass through the data; using the $QR$ factorization would be more stable, and could also be done in one pass. LARS for nonlinear regression requires multiple passes through the data for each step, hence speed becomes much more of an issue.

### 2.2.2. Ridge Regression

The ad-hoc nature and instability of variable selection methods has led to other approaches. Ridge regression (Miller, 2002; Draper and Smith, 1998), includes all predictors, but with typically smaller coefficients than they would have under ordinary least squares. The coefficients minimize a penalized sum of squares,

$$\|\mathbf{Y} - \mathbf{X}\boldsymbol{\beta}\|_2^2 + \theta \sum_{j=1}^{p} \beta_j^2. \tag{2}$$

where $\theta$ is a positive scalar; $\theta = 0$ corresponds to ordinary least-squares regression. In practice no penalty is applied to the intercept, and variables are scaled to variance 1 so that the penalty is invariant to the scale of the original data.

Figure 1 shows the coefficients for ridge regression graphically as a function of $\theta$; these shrink as $\theta$ increases. Variables most correlated with other variables are affected most, e.g. S1 and S2 have correlation 0.90.

Note that as $\theta$ increases, the coefficients approach but do not equal zero. Hence, no variable is ever excluded from the model (except when coefficients cross zero for smaller values of $\theta$).

In contrast, the use of an $\ell_1$ penalty does reduce terms to zero. This yields LASSO, which we consider next.

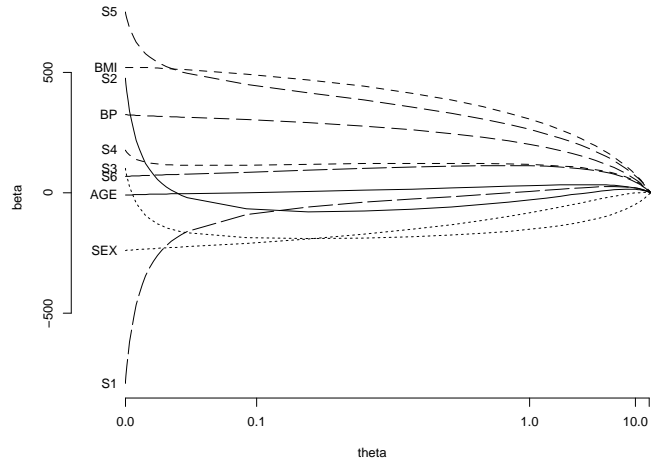

FIG 1. *Coefficients for ridge regression (standardized variables)*

### 2.2.3. *LASSO*

Tibshirani (1996) proposed minimizing the residual sum of squares, subject to a constraint on the sum of absolute values of the regression coefficients, $\sum_{j=1}^{p} |\beta_j| \leq t$. This is equivalent to minimizing the sums of squares of residuals plus an $\ell_1$ penalty on the regression coefficients,

$$\|\mathbf{Y} - \mathbf{X}\boldsymbol{\beta}\|_2^2 + \theta \sum_{j=1}^{p} |\beta_j|. \tag{3}$$

A similar formulation was proposed by Chen *et al.* (1998) under the name *basis pursuit*, for denoising using overcomplete wavelet dictionaries (this corresponds to $p > n$).

Figure 2 shows the resulting coefficients. For comparison, the right panel shows the coefficients from ridge regression, plotted on the same scale. To the right, where the penalties are small, the two procedures give close to the same results. More interesting is what happens starting from the left, as all coefficients start at zero and penalties are relaxed. For ridge regression all coefficients immediately become nonzero. For LASSO, coefficients become nonzero one at a time. Hence the $\ell_1$ penalty results in variable selection, as variables with coefficients of zero are effectively omitted from the model.

Another important difference occurs for the predictors that are most significant. Whereas an $\ell_2$ penalty $\theta \sum \beta_j^2$ pushes $\beta_j$ toward zero with a force proportional to the value of the coefficient, an $\ell_1$ penalty $\theta \sum |\beta_j|$ exerts the same force on all nonzero coefficients. Hence for variables that are most valuable, that clearly should be in the model and where shrinkage toward zero is less desirable, an $\ell_1$ penalty shrinks less. This is important for providing accurate predictions of future values.

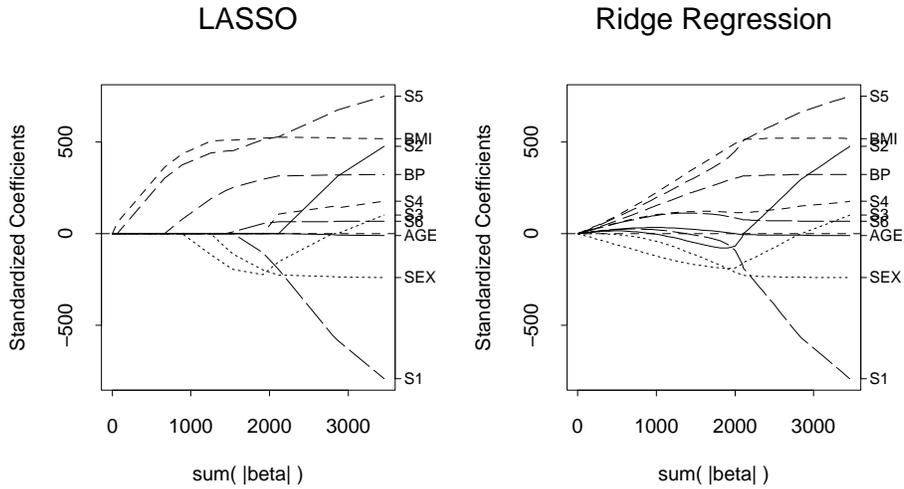

Fig 2. *Coefficients for LASSO and Ridge Regression ($\ell_1$ and $\ell_2$ penalties).*

In this case, BMI (body mass index) and S5 (a blood serum measurement) appear to be most important, followed by BP (blood pressure), S3, Sex, S6, S1, S4, S2, and Age. Some curious features are apparent. S1 and S2 enter the model relatively late, but when they do their coefficients grow rapidly, in opposite directions. These two variables have strong positive correlation, so these terms largely cancel out, with little effect on predictions for the observed values. The collinearity between these two variables has a number of undesirable consequences—relatively small changes in the data can have strong effects on the coefficients, the coefficients are unstable, predictions for new data may be unstable, particularly if the new data do not follow the same relationship between S1 and S2 found in the training data, and the calculation of coefficients may be numerically inaccurate. Also, the S3 coefficient changes direction when S4 enters the model, ultimately changing sign. This is due to high (negative) correlation between S3 and S4.

### 2.2.4. Forward Stagewise

Another procedure, forward stagewise regression, appears to be very different from LASSO, but turns out to have similar behavior.

This procedure is motivated by a desire to mitigate the negative effects of the greedy behavior of stepwise regression. In stepwise regression, the most useful predictor is added to the model at each step, and the coefficient jumps from zero to the the least-squares value.

Forward stagewise picks the same first variable as forward stepwise, but changes the corresponding coefficient only a small amount. It then picks the variable with highest correlation with the current residuals (possibly the same variable as in the previous step), and takes a small step for that variable, and continues in this fashion.

Where one variable has a clear initial advantage over other variables there will be a number of steps taken for that variable. Subsequently, once a number of variables are in the model, the procedure tends to alternate between them. The resulting coefficients are more stable than those for stepwise.

Curiously, an idealized version of forward stagewise regression (with the step size tending toward zero) has very similar behavior to LASSO despite the apparent differences. In the diabetes example, the two methods give identical results until the eighth variable enters, after which there are small differences (Efron *et al.*, 2004).

There are also strong connections between forward stagewise regression and the boosting algorithm popular in machine learning (Efron *et al.* 2004; Hastie *et al.* 2001). The difference is not in the fitting method, but rather in the predictors used; in stagewise the predictors are typically determined in advance, while in boosting the next variable is typically determined on the fly.

### 2.2.5. Least Angle Regression

Least angle regression (Efron *et al.*, 2004) can be viewed as a version of stagewise that uses mathematical formulas to accelerate the computations. Rather than taking many tiny steps with the first variable, the appropriate number of steps is determined algebraically, until the second variable begins to enter the model. Then, rather than taking alternating steps between those two variables until a third variable enters the model, the method jumps right to the appropriate spot. Figure 3 shows this process in the case of 2 predictor variables, for linear regression.

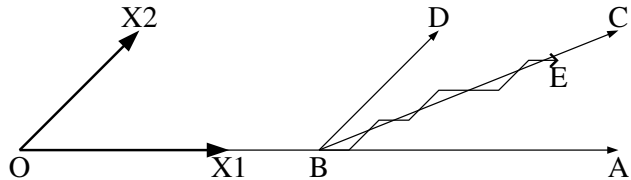

FIG 3. *The LAR algorithm in the case of 2 predictors. O is the prediction based solely on an intercept. $C = \hat{Y} = \hat{\beta}_1 X_1 + \hat{\beta}_2 X_2$ is the ordinary least-squares fit, the projection of Y onto the subspace spanned by $X_1$ and $X_2$. A is the forward stepwise fit after one step; the second step proceeds to C. Stagewise takes a number of tiny steps from O to B, then takes steps alternating between the $X_1$ and $X_2$ directions, eventually reaching E; if allowed to continue it would reach C. LAR jumps from O to B in one step, where B is the point such that BC bisects the angle ABD. At the second step it jumps to C. LASSO follows a path from O to B, then from B to C. Here LAR agrees with LASSO and stagewise (as the step size → 0 for stagewise). In higher dimensions additional conditions are needed for exact agreement to hold.*

The first variable chosen is the one that has the smallest angle between the variable and the response variable; in Figure 3 the angle $COX_1$ is smaller than $COX_2$. We proceed in that direction as long as the angle between that predictor and the vector of residuals $Y - \gamma X_1$ is smaller than the angle between other predictors and the residuals. Eventually the angle for another variable will equal this angle (once we reach point $B$ in Figure 3), at which point we begin moving toward the direction of the least-squares fit based on both variables. In higher dimensions we will reach the point at which a third variable has an equal angle, and joins the model, etc.

Expressed another way, the (absolute value of the) correlation between the residuals and the first predictor is greater than the (absolute) correlation for other predictors. As $\gamma$ increases, another variable will eventually have a correlation with the residuals equaling that of the active variable, and join the model as a second active variable. In higher dimensions additional variables will eventually join the model, when the correlation between all active variables and the residuals drops to the levels of the additional variables.

**Three remarkable properties of LAR**   There are three remarkable things about LAR. First is the speed: Efron *et al.* (2004) note that "The entire sequence of LARS steps with $p < n$ variables requires $O(p^3 + np^2)$ computations — the cost of a least squares fit on $p$ variables."

Second is that the basic LAR algorithm, based on the geometry of angle bisection, can be used to efficiently fit LASSO and stagewise models, with certain modifications in higher dimensions (Efron *et al.*, 2004). This provides a fast and relatively simple way to fit LASSO and stagewise models.

Madigan and Ridgeway (2004) comments that LASSO has had little impact on statistical practice, due to the inefficiency of the original LASSO and complexity of more recent algorithms (Osborne *et al.*, 2000a); they add that this "efficient, simple algorithm for the LASSO as well as algorithms for stagewise regression and the new least angle regression" are "an important contribution to statistical computing".

Third is the availability of a simple $C_p$ statistic for choosing the number of steps,

$$C_p = (1/\hat{\sigma}^2) \sum_{i=1}^{n} (y_i - \hat{y}_i)^2 - n + 2k \tag{4}$$

where $k$ is the number of steps and $\hat{\sigma}^2$ is the estimated residual variance (estimated from the saturated model, assuming that $n > p$). This is based on Theorem 3 in Efron *et al.* (2004), which indicates that after $k$ steps of LAR the degrees of freedom $\sum_{i=1}^{n} \operatorname{cov}(\hat{\mu}_i, Y_i)/\sigma^2$ is approximately $k$. This provides a simple stopping rule, to stop after the number of steps $k$ that minimizes the $C_p$ statistic.

Zou *et al.* (2007) extend that result to LASSO, showing an unbiased relationship between the number of terms in the model and degrees of freedom, and discuss $C_p$, AIC and BIC criterion for model selection.

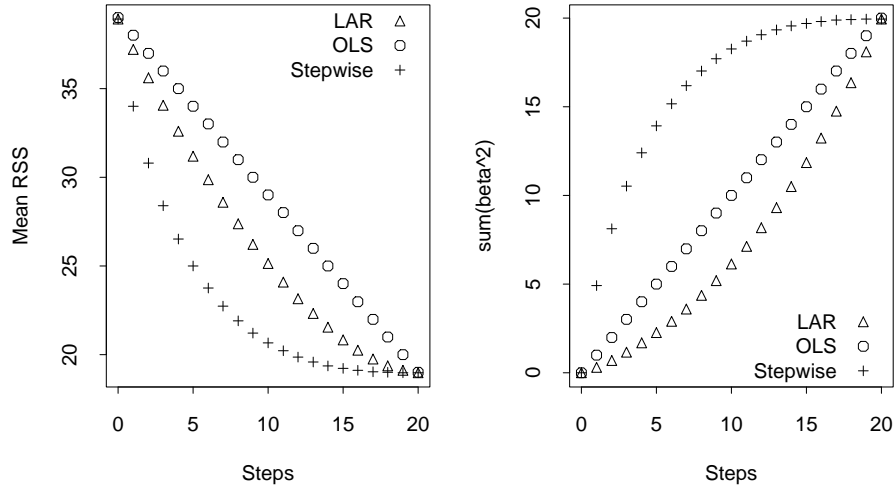

FIG 4. *Effect of LAR steps on residual variance and prediction error. The left panel shows the residual sum of squares for LAR, ordinary least-squares with fixed predictor order, and stepwise regression. The right panel shows $\sum_{i=1}^{p} \beta_j^2$; this measures how much less accurate predictions are than for the true model. The figures are based on a simulation with 10,000 replications, with $n = 40$, $p = 20$, orthogonal predictors with norm 1, $\beta_j = 1 \forall j$, and residual variance 1.*

The promise of a fast effective way of choosing the tuning parameter, based on $C_p$, AIC or BIC, is important in practice. While figures such as Figure 2 are attractive, they become unwieldy in high dimensions. In any case, for prediction one must ultimately choose a single value of the penalty parameter.

Still, there are some questions about this $C_p$ statistic (Ishwaran 2004; Loubes and Massart 2004; Madigan and Ridgeway 2004; Stine 2004), and some suggest other selection criteria, especially cross-validation.

Cross-validation is slow. Still, a fast approximation for the tuning parameter could speed up cross-validation. For example, suppose there are 1000 predictors, and $C_p$ suggests that the optimal number to include in a model is 20; then when doing cross-validation one might stop after say 40 steps in every iteration, rather 1000.

Note that there are different definitions of degrees of freedom, and the one used here is appropriate for $C_p$ statistics, but that $k$ does not measure other kinds of degrees of freedom. In particular, neither the average drop in residual squared error, nor the expected prediction error are linear in $k$ (under the null hypothesis that $\beta_j = 0$ for all $j$). Figure 4 shows the behavior of those quantities. In the left panel we see that the residual sums of squares drop more quickly for LAR than for ordinary least squares (OLS) with fixed prediction order, suggesting that by one measure, the effective degrees of freedom is greater than $k$. In the right panel, the sums of squares of coefficients measures how much worse predictions are than using the true parameters $\beta_j = 0$; here LAR increases more slowly than for OLS, suggesting effective degrees of freedom less than $k$. These two effects balance out for the $C_p$ statistic.

In contrast, stepwise regression has effective degrees of freedom greater than the number of steps; it overfits when there is no true signal, and prediction errors suffer.

These results are encouraging. It appears that LAR fits the data more closely than OLS, with a smaller penalty in prediction errors. While in this example there is only noise and no signal, it suggests that LAR may have relatively high sensitivity to signal and low sensitivity to noise.

### 2.2.6. Comparing LAR, LASSO and Stagewise

In general in higher dimensions native LAR and the least angle implementation of LASSO and stagewise give results that are similar but not identical. When they differ, LAR has a speed advantage, because LAR variables are added to the model, never removed. Hence it will reach the full least-squares solution, using all variables, in $p$ steps. For LASSO, and to a greater extent for stagewise, variables can leave the model, and possibly re-enter later, multiple times. Hence they may take more than $p$ steps to reach the full model (if $n > p$). Efron *et al.* (2004) test the three procedures for the diabetes data using a quadratic model, consisting of the 10 main effects, 45 two-way interactions, and 9 squares (excluding the binary variable Sex). LAR takes 64 steps to reach the full model, LASSO takes 103, and stagewise takes 255. Even in other situations, when stopping short of the saturated model, LAR has a speed advantage.

The three methods have interesting derivations. LASSO is regression with an $\ell_1$ penalty, a relatively simple concept; this is also known as a form of regularization in the machine learning community. Stagewise is closely related to boosting, or "slow learning" in machine learning (Efron *et al.*, 2004; Hastie *et al.*, 2007). LAR has a simpler interpretation than the original derivation; it can be viewed as a variation of Newton's method (Hesterberg and Fraley 2006a, 2006b), which makes it easier to extend to some nonlinear models such as generalized linear models (Rosset and Zhu, 2004).

## 3. LARS Extensions

In this section we review extensions to LARS and other contributions described in the literature. We introduce LARS extensions that account for specific structures in variables in Section 3.1, extensions to nonlinear models in Section 3.2, extensions in other settings in Section 3.3, and computational issues in Section 3.4.

Ridge regression and LASSO optimize a criterion that includes a penalty term. A number of authors develop other penalty approaches, including SCAD (Fan and Li, 2001), adaptive LASSO (Zou, 2006), relaxed LASSO (Meinshausen, 2007), and the Dantzig selector (Candes and Tao, 2007). Some of these may be considered as alternatives rather than extensions to LARS, so we defer this discussion until Section 4.

### 3.1. Exploiting Additional Structure

Some kinds of data have structure in the predictor variables—they may be ordered in some meaningful way (such as measurements based on intensity at successive wavelengths of light in proteomics) or come in groups, either known (such as groups of dummy variables for a factor) or unknown (such as related genes in microarray analysis). There may be order restrictions (such as main effects before interactions).

When there is a group of strongly correlated predictors, LASSO tends to select only one predictor from the group, but we may prefer to select the whole group. For a sequence of ordered predictors, we may want the differences between successive coefficients to be small.

### 3.1.1. Ordered Predictors

Tibshirani *et al.* (2005) propose the *fused LASSO* for a sequence of predictors. This uses a combination of an $\ell_1$ penalty on coefficients and an $\ell_1$ penalty on the difference between adjacent coefficients:

$$\|\mathbf{Y} - \mathbf{X}\boldsymbol{\beta}\|_2^2 + \theta_1 \sum_{j=1}^{p} |\beta_j| + \theta_2 \sum_{j=2}^{p} |\beta_j - \beta_{j-1}|.$$

This differs from LASSO in that the additional $\ell_1$ penalty on the difference between successive coefficients encourages the coefficient profiles $\beta_j$ (a function of $j$) to be locally flat. The fused LASSO is useful for problems such as the analysis of proteomics data, where there is a natural ordering of the predictors (e.g. measurements on different wavelengths) and coefficients for nearby predictors should normally be similar; it tends to give locally-constant coefficients. Estimates can be obtained via a quadratic programming approach for a fixed pair $(\theta_1, \theta_2)$, or by pathwise coordinate optimization (Friedman *et al.*, 2007a).

### 3.1.2. Unknown Predictor Groups

Zou and Hastie (2005b) propose the *elastic net* [1] for applications with unknown groups of predictors. It involves both the $\ell_1$ penalty from LASSO and the $\ell_2$ penalty from ridge regression:

$$\|\mathbf{Y} - \mathbf{X}\boldsymbol{\beta}\|_2^2 + \theta_1 \sum_{j=1}^{p} |\beta_j| + \theta_2 \sum_{j=1}^{p} \beta_j^2 \qquad (5)$$

They show that strictly convex penalty functions have a grouping effect, while the LASSO $\ell_1$ penalty does not. A bridge regression (Frank and Friedman, 1993) $\ell_q$ norm penalty with $1 < q < 2$ is strictly convex and has a grouping effect, but does not produce a sparse solution (Fan and Li, 2001). This motivates

---

[1] R package `elasticnet` is available.

Zou and Hastie (2005b) to use the elastic net penalty (5), which is strictly convex when $\theta_2 > 0$, and can also produce sparse solutions. The elastic net is useful in the analysis of microarray data, as it tends to bring related genes into the model as a group. It also appears to give better predictions than LASSO when predictors are correlated. In high dimensional settings ($p \gg n$) elastic net allows selecting more than $n$ predictors, while LASSO does not. Solutions can be computed efficiently using an algorithm based on LARS; for given $\theta_2$, formula (5) can be interpreted as a LASSO problem.

### 3.1.3. Known Predictor Groups

In some cases it is appropriate to select or drop a group of variables simultaneously, for example a set of dummy variables that represent a multi-level factor. Similarly, a set of basis functions for a polynomial or spline fit should be treated as a group.

Yuan and Lin (2006) propose *group LASSO* to handle groups of predictors (see also (Bakin, 1999)). Suppose the $p$ predictors are divided into $J$ groups of sizes $p_1, \ldots, p_J$, and let $\boldsymbol{\beta}_j$ be the corresponding sub-vectors of $\boldsymbol{\beta}$. Group LASSO minimizes

$$\|\mathbf{Y} - \mathbf{X}\boldsymbol{\beta}\|_2^2 + \theta \sum_{j=1}^{p} \|\boldsymbol{\beta}_j\|_{K_j}, \qquad (6)$$

where $\|\eta\|_K = (\eta^T K \eta)^{1/2}$ is the elliptical norm determined by a positive definite matrix $K$. This includes LASSO as a special case, with $p_j = 1$ for all $j$ and each $K_j$ the one-dimensional identity matrix. Yuan and Lin (2006) use $K_j = p_j I_{p_j}$, where $I_{p_j}$ is the $p_j$-dimensional identity matrix. The modified penalty in (6) encourages sparsity in the number of groups included, rather than the number of variables.

Lin and Zhang (2006) let the groups of predictors correspond to sets of basis functions for smoothing splines, in which the penalty $\|\beta_j\|_{K_j}$ would give the square-root of the integrated squared second derivative of a spline function (a linear combination of the basis functions). Their resulting *COSSO* (COmponent Selection and Smoothing Operator) is an alternative to MARS (Friedman 1991).

Yuan and Lin (2006) note that group LASSO does not have piecewise linear solution paths, and define a *group LARS* that does. Group LARS replaces the correlation criterion in the original LARS with the average squared correlation between a group of variables and the current residual. A group of variables that has the highest average squared correlation with the residual is added to the active set. Park and Hastie (2006b) modify group LARS, replacing the average squared correlation with the average absolute correlation to prevent selecting a large group with only few of its components being correlated with the residuals.

The *Composite Absolute Penalties* (CAP) approach, proposed by Zhao *et al.* (2008), is similar to group LASSO but uses $\ell_{\gamma_j}$-norm instead of $\ell_2$-norm, and the equivalent of an $\ell_{\gamma_0}$ norm for combining group penalties:

$$\|\mathbf{Y} - \mathbf{X}\boldsymbol{\beta}\|_2^2 + \theta \sum_{j=1}^{J} (\|\boldsymbol{\beta}_j\|_{\gamma_j})^{\gamma_0} \qquad (7)$$

where $\gamma_j > 1$ for grouped variable selection. For example, when $\gamma_j = \infty$, the coefficients in the $j^{\text{th}}$ group are encouraged to be of equal size, while $\gamma_j = 2$ does not imply any information but the grouping information.

An obvious generalization that could apply to many of the methods, both in grouped and ungrouped settings, is to include constant factors in the penalties for variables or groups to penalize different terms different amounts. Yuan and Lin (2006) include constant terms $p_j$ depending on degrees of freedom—terms with more degrees of freedom are penalized more. Similar constants could be used to reflect the desirability of penalizing different terms differently. For example, some terms known from previous experience to be important could be left unpenalized or penalized using a small coefficient, while a larger number of terms being screened as possible contributors could be assigned higher penalties. Main effects could be penalized by small amounts and higher-order interactions penalized more.

### 3.1.4. Order Restrictions

Besides group structure, we may want to incorporate order restrictions in variable selection procedures. For example, a higher order term (e.g. an interaction term $X_1 X_2$) should be selected only when the corresponding lower order terms (e.g. main effects $X_1$ and $X_2$) are present in the model. This is the marginality principle in linear models (McCullagh and Nelder, 1989) and heredity principle in design of experiments (Hamada and Wu, 1992). Although it is not a strict rule, it is usually better to enforce order restriction, because it helps the resulting models to be invariant to scaling and transformation of predictors.

Efron *et al.* (2004) suggest a two-step procedure to enforce order restrictions: first apply LARS only to main effects, and then to possible interactions between the main effects selected from the first step. Turlach (2004) shows that the two-step procedure may miss important main effects at the first step in some nontrivial cases and proposes an extended version of LARS: when the $j^{\text{th}}$ variable has the highest correlation with the residual, that variable and a set of variables on which it depends enter the model together. Yuan *et al.* (2007) propose a similar extension to LARS that accounts for the number of variables that enter the model together: they look at the scaled correlations between the response and the linear space spanned by the set of variables that should be selected together. Choi and Zhu (2006) discuss re-parameterizing the interaction coefficients to incorporate order restrictions, and the CAP approach (7) of Zhao *et al.* (2008) can be used for the same purpose by assigning overlapping groups (e.g. groups for each main effect and another that includes interactions and all main effects).

There is another type of order restriction called weak heredity or marginality principle: a higher order term can be selected only when at least one of the corresponding lower order terms is present in the model. Yuan *et al.* (2007) extend LARS to this case by looking at the scaled correlations between the response and the linear space spanned by each eligible set of predictors; in contrast to the strong heredity case, the combination of an interaction and just one (rather than both) of the corresponding main effects would be eligible.

### 3.1.5. Time Series and Multiresponse Data

The fused LASSO introduced in Section 3.1.1 is for problems with a sequence of ordered predictors. Some problems, however, contain natural orderings in response variables as well. A good example would be time-course data, in which the data consist of multiple observations over time; either responses or predictors, or both, could vary over time. For such cases, we could simply fit a model at each time point, but it would be more efficient to combine the information from the entire dataset. As an illustration, consider linear regression with multiple responses at $N$ different time points $t_1, \ldots, t_N$ and fixed predictors $X$:

$$\mathbf{Y}(t_r) = \mathbf{X}\boldsymbol{\beta}(t_r) + \boldsymbol{\epsilon}(t_r), \quad r = 1, \ldots, N. \tag{8}$$

where $\mathbf{Y}(t_r) \in \mathbb{R}^n$, $\boldsymbol{\beta}(t_r) \in \mathbb{R}^p$, $\boldsymbol{\epsilon}(t_r) \in \mathbb{R}^n$, and $\mathbf{X}$ is a $n \times p$ design matrix. By assuming that adjacent time points are related and similar, we could apply the fused LASSO to this problem by penalizing the difference between the coefficients of successive time points, $|\beta_j(t_r) - \beta_j(t_{r-1})|$. But it could be challenging to simultaneously fit a model with all $Np$ parameters when the number of time points $N$ is large.

Meier and Bühlmann (2007) propose *smoothed LASSO* to solve this problem. They assume that adjacent time points are more related than distant time points, and incorporate the information from different time points by applying weights $w(\cdot, t_r)$ satisfying $\sum_{s=1}^{N} w(t_s, t_r) = 1$ in the criterion below for parameter estimation at time-point $t_r$:

$$\sum_{s=1}^{N} w(t_s, t_r) \|\mathbf{Y}(t_s) - \mathbf{X}\boldsymbol{\beta}(t_r)\|_2^2 + \theta \sum_{j=1}^{p} |\beta_j(t_r)|. \tag{9}$$

The weights $w(\cdot, t_r)$ should have larger values at the time points near $t_r$ so that the resulting estimates can reflect more information from neighboring points. Problem (9) can be solved as an ordinary LASSO problem by using the smoothed response $\tilde{\mathbf{Y}}(t_r) = \sum_{s=1}^{N} w(t_s, t_r)\mathbf{Y}(t_s)$.

Turlach *et al.* (2005) and Similä and Tikka (2006) also address the multiple response problem with different approaches. Turlach *et al.* (2005) extend LASSO to select a common subset of predictors for predicting multiple response variables using the following criterion:

$$\sum_{r=1}^{N} \|\mathbf{Y}(t_r) - \mathbf{X}\boldsymbol{\beta}(t_r)\|_2^2 + \theta \sum_{j=1}^{p} \max_{r=1,\ldots,N} |\beta_j(t_r)|. \tag{10}$$

We note that this is equivalent to a special case (with $\gamma_j = \infty$) of the CAP approach (Zhao *et al.*, 2008) that was introduced in Section 3.1.3 for grouped variable selection. On the other hand, Similä and Tikka (2006) extend the LARS algorithm by defining a new correlation criterion between the residuals and the predictor, $\|(\mathbf{Y} - \hat{\mathbf{Y}})^T x_j\|_\gamma$ ($\gamma = 1, 2, \infty$) where $\mathbf{Y} = (y(t_1), \dots, y(t_N))$ is an $n \times N$ matrix. They note that their method is very similar to group LARS (Yuan and Lin, 2006) when $\gamma = 2$. Both of their procedures differ from the smoothed LASSO in that all coefficients corresponding to one predictor are estimated as either zero or nonzero as a group — if a predictor is selected, its coefficients at different time points are all nonzero, in contrast to the smoothed LASSO which may have different nonzero coefficients at different times.

In (8), the predictors $X$ are the same for different time points, but in some applications both $X$ and $y$ can vary over time. Balakrishnan and Madigan (2007) combine ideas from group LASSO (Yuan and Lin, 2006) and fused LASSO (Tibshirani *et al.*, 2005), aiming to select important groups of correlated time-series predictors. Wang *et al.* (2007b) consider autoregressive error models that involve two kinds of coefficients, regression coefficients and autoregression coefficients. By applying two separate $\ell_1$ penalties to regression coefficients and autoregression coefficients, they achieve a sparse model that includes both important predictors and autoregression terms.

### *3.2. Nonlinear models*

The original LARS method is for linear regression:

$$E(Y|\mathbf{X} = \mathbf{x}) = f(\mathbf{x}) = \beta_0 + \beta_1 x_1 + \dots + \beta_p x_p, \tag{11}$$

where the regression function $f(\mathbf{x})$ has a linear relationship to the predictors $x_1, \dots, x_p$ through the coefficients $\beta_1, \dots, \beta_p$. The problem can also be viewed as the minimization of a sum-of-squares criterion

$$\min_{\boldsymbol{\beta}} \|\mathbf{Y} - \mathbf{X}\boldsymbol{\beta}\|_2^2,$$

with added variable or model selection considerations. The LASSO extension gives an efficient solution for the case of an $\ell_1$ penalty term on regression coefficients:

$$\min_{\boldsymbol{\beta}} \|Y - \mathbf{X}\boldsymbol{\beta}\|_2^2 + \theta \sum_{j=1}^{p} |\beta_j|. \tag{12}$$

The number of solutions to (12) is finite for $\theta \in [0, \infty)$, and predictor selection is automatic since the solutions vary in the number and location of nonzero coefficients.

The original LARS methods apply to quite general models of the form

$$E(Y|\mathbf{X} = \mathbf{x}) = f(\mathbf{x}) = \beta_0 + \beta_1 \phi_1(\mathbf{x}) + \dots + \beta_M \phi_M(\mathbf{x}), \tag{13}$$

where $\phi_m$ are (nonlinear) functions of the original predictors $\mathbf{X}$. The $\phi_m$ could, for example, include higher-order terms and interactions such as $x_i^2$ or $x_i x_j$, nonlinear transformations such as $\log(x_i)$, piecewise polynomials, splines and kernels.

The use of nonlinear basis functions $\phi_j(\mathbf{x})$ allows the use of linear methods for fitting nonlinear relationships between $y$ and the predictors $x_j$. As long as the $\phi_m$ are predetermined, the fundamental structure of the problem is linear and the original LARS methods are applicable. For example, Avalos *et al.* (2007) consider additive models where each additive component $\phi_j(\mathbf{x}) = \phi_j(x_j)$ is fitted by cubic splines. They discuss the extension of LASSO to those models by imposing the $\ell_1$ penalty on the coefficients of the linear part to get a sparse model. A drawback is that the resulting model may not obey order restrictions; for example it may drop a linear term while keeping the corresponding higher order terms.

Another example is kernel regression, in which $\phi_m(\mathbf{x}) = K_\lambda(x_m, \mathbf{x})$ ($m = 1, \ldots, n$), where $K$ is a kernel function belonging to a reproducing kernel Hilbert space (RKHS), and $\lambda$ is a hyperparameter that regulates the scale of the kernel function $K$. By imposing an $\ell_1$ penalty on the coefficients with the squared error loss function, the resulting model has a sparse representation based on a smaller number of kernels so that predictions can be computed more efficiently. Wang *et al.* (2007a) discuss a path-following algorithm based on LARS to fit solutions to this $\ell_1$ regularized kernel regression model, as well as a separate path-following algorithm for estimating the optimal kernel hyperparameter $\lambda$. Guigue *et al.* (2006) and Gunn and Kandola (2002) consider LASSO extensions to more flexible kernel regression models, in which each kernel function $K_\lambda(x_m, \cdot)$ is replaced by a weighted sum of multiple kernels.

More generally, the sum-of-squares loss function in (12) can be replaced by a more general convex loss function $\mathcal{L}$,

$$\min_{\boldsymbol{\beta}} \mathcal{L}(y, \phi(\mathbf{x})\boldsymbol{\beta}) + \theta \sum_{j=1}^{n} |\beta_j|, \qquad (14)$$

although solution strategies become more complicated. Rosset and Zhu (2007) extend the LARS-LASSO algorithm to use Huber's loss function by specifying modifications when the solution path hits the knots between the linear part and quadratic part. Huber's loss is also considered in Roth (2004) for $\ell_1$ regularized kernel regression based on iteratively reweighted least squares (IRLS). When $\mathcal{L}$ is $\epsilon$-insensitive loss, $\mathcal{L}_\epsilon(y, \hat{y}) = \sum_{i=1}^{n} \max(0, |y_i - \hat{y}_i| - \epsilon)$, the problem becomes an $\ell_1$ regularized Support Vector Machine (SVM). Path-following algorithms for this problem are discussed in Zhu *et al.* (2003) and Hastie *et al.* (2004).

In several important applications, including generalized linear models and Cox proportional hazards models, some function of the regression function $f(x)$ is linearly associated with the parameters $\beta$:

$$g(E(Y|\mathbf{X} = \mathbf{x})) = g(f(\mathbf{x})) = \beta_0 + \beta_1 x_1 + \ldots + \beta_p x_p. \qquad (15)$$

Several authors discuss extensions of LARS to these models: generalized linear models (Lokhorst, 1999; Roth, 2004; Madigan and Ridgeway, 2004; Rosset, 2005; Park and Hastie, 2007; Keerthi and Shevade, 2007) and Cox regression (Tibshirani, 1997; Gui and Li, 2005; Park and Hastie, 2007). [2]

Some authors focus on the special case of a binary response (logistic regression). The function $g$ in (15) has a parametric form and is linearly related to predictors **x**. Zhang *et al.* (2004) consider a nonparametric framework called *Smoothing Spline ANOVA* and extend LASSO by using the penalized negative Bernoulli log-likelihood with an $\ell_1$ penalty on the coefficients of the basis functions. Shi *et al.* (2008) consider a two-step procedure to efficiently explore potential high order interaction patterns for predicting the binary response in high dimensional data where the number of predictors is very large. They first focus on to binary (or dichotomized) predictors, and impose an $\ell_1$ penalty on the coefficients of the basis functions for main effects and higher-order interactions of those binary predictors to achieve a sparse representation. They then use only the selected basis functions to fit a final linear logistic model.

The preceding paragraphs discuss applications with particular loss functions; some authors propose general strategies for LASSO problems with general convex loss functions. Rosset and Zhu (2007) discuss conditions under which coefficient paths are piecewise linear. Rosset (2005) discusses a method for tracking curved coefficient paths for which the computational requirements severely limit its suitability for large problems. Kim *et al.* (2005b) propose a gradient approach[3] that is particularly useful for high dimensions due to computationally affordability; it requires only a univariate optimization at each iteration, and its convergence rate is independent of the data dimension. Wang and Leng (2007) suggest using approximations to loss functions that are quadratic functions of the coefficients, so that solutions can then be computed using the LARS algorithm.

Boosting is another technique that can be used to approximately fit $\ell_1$ regularized models. Efron *et al.* (2004) showed that forward stagewise regression can be viewed as a version of boosting for linear regression with the squared error loss, producing a similar result to LASSO when the step size approaches zero. For general loss functions, Zhao and Yu (2007) approximate the LASSO solution path by incorporating forward stagewise fitting and backward steps. Friedman (2006) discusses a gradient boosting based method that can be applied to general penalty functions as well as general loss functions.

Some of the approaches introduced in Section 3.1 for grouped and ordered predictors have also been extended to nonlinear models. Park and Hastie (2007) extend a path-following algorithm for elastic net to generalized linear models for a fixed $\theta_2$ in (5). They note that adding an $\ell_2$ penalty is especially useful for logistic regression since it prevents $\|\beta\|_1$ from growing to infinity as the regularization parameter $\theta$ decreases to zero, a common problem that arises in $\ell_1$ fitting to separable data. Park and Hastie (2006b) propose a path-following algorithm for group LASSO in exponential family models. Kim *et al.* (2006) use a gradient projection method to extend group LASSO to general loss functions, and Meier *et al.* (2008)[4] discuss an algorithm for group LASSO for logistic regression models.

---

[2] S-PLUS and R packages `glmpath` and `glars` are available, for both GLMs and Cox regression.

[3] R-package `glasso` is available.

### *3.3. Other Applications*

**Robust Regression**  Rosset and Zhu (2007) and Owen (2006) extend LASSO by replacing the squared error loss by Huber's loss. In the linear regression case this also yields piecewise-linear solution paths, allowing for fast solutions. Khan *et al.* (2007) extend LAR by replacing correlations with robust correlation estimates.

**Subset of Observations**  LARS can be used for choosing an important subset of observations as well as for selecting a subset of variables. Silva *et al.* (2005) apply LARS for selecting a representative subset of the data for use as landmarks to reduce computational expense in nonlinear manifold models.

**Principal Component and Discriminant Analysis**  Jolliffe *et al.* (2003) and Zou *et al.* (2006) apply $\ell_1$ penalties to get sparse loadings in principal components. Trendafilov and Joilliffe (2007) discuss $\ell_1$ penalties in linear discriminant analysis.

**Gaussian Graphical Models**  A number of authors discuss using $\ell_1$ penalties to estimate a sparse inverse covariance matrix (or a sparse graphical model). Meinshausen and Bühlmann (2006) fit a LASSO model to each variable, using the others as predictors, then set the $ij$ term of the inverse covariance matrix to zero if the coefficient of $X_j$ for predicting $X_i$ is zero, or the converse. Many authors (Yuan, 2008; Banerjee *et al.*, 2008; Dahl *et al.*, 2008; Yuan and Lin, 2007a; Friedman *et al.*, 2007b) discuss efficient methods for optimizing the $\ell_1$-penalized likelihood, using interior-point or blockwise coordinate-descent approaches. This work has yet to be extended to handle nonlinear relationships between variables, such as (13).

### *3.4. Computational Issues*

There are three primary computation issues: speed, memory usage, and numerical accuracy.

The original LAR algorithm for linear regression as described in Efron *et al.* (2004) and implemented in Efron and Hastie (2003)[5] is remarkably fast and memory efficient in the $p < n$ case, as noted in Section 2.1. Minor modifications allow computing the LASSO and forward stagewise cases. However, the implementations use cross-product matrices, which are notorious for numerical inaccuracy with highly correlated predictors.

---

[4] R-package `grplasso` is available.
[5] S-PLUS and R package `lars` is available.

Fraley and Hesterberg (2007) (see also Hesterberg and Fraley 2006a,b) develop LARS implementations based on $QR$ decompositions [6]. This reduces the roundoff error by a factor equal to the condition number of $X$ relative to the original algorithm. One variation uses only a single pass through the data for an initial factorization, after which it requires storage $O(p^2)$, independent of $n$; in contrast the original LARS implementation is intended for in-memory datasets, and makes multiple passes through the data.

Fu (1998) proposes a shooting algorithm to solve LASSO for a specified value of the penalty parameter $\theta$. The algorithm is a special case of a coordinate descent method that cycles through the coordinates, optimizing the current one and keeping the remaining coordinates fixed. Using a (predefined) grid of penalty parameters, the coefficient paths can be computed efficiently, especially in very high-dimensional settings, by making use of the preceding solution as starting values.

Other coordinate-wise optimization techniques have shown their success with other penalty types and also for nonlinear models (Genkin *et al.* 2007; Yuan and Lin 2006; Meier *et al.* 2008; Friedman *et al.* 2007a,b).

Osborne *et al.* (2000a) propose a descent algorithm for a LASSO problem with a specified value of the penalty parameter $\theta$, as well as a homotopy method for the piecewise linear solution path in the linear regression case that is related to the LAR implementation of Efron *et al.* (2004). In Osborne *et al.* (2000b), an algorithm based on LASSO and its dual is proposed that yields new insights and an improved method for estimating standard errors of regression parameters.

**Nonlinear regression**  In the linear regression case the solution path is piecewise linear, and each step direction and jump size can be computed in closed-form solution. In the nonlinear case paths are curved, so that iterative methods are needed for computing and updating directions and determining the ends of each curve, requiring multiple passes through the data. Hence the algorithms are much slower than in the linear case.

## 4. Theoretical Properties and Alternative Regularization Approaches

In this section we discuss some theoretical properties of LASSO, and illustrate how some alternative regularization approaches address the drawbacks of LASSO.

### 4.1. Criteria

It is important to distinguish between the goals of prediction accuracy and variable selection. If the main interest is in finding an interpretable model or in identifying the "true" underlying model as closely as possible, prediction accuracy is of secondary importance. An example would be network modeling in biology. On the other hand, if prediction is the focus of interest, it is usually acceptable for the selected model to contain some extra variables, as long as the coefficients of those variables are small.

---

[6]S-PLUS and R package `sclars` is available.

### 4.1.1. The Prediction Problem

Greenshtein and Ritov (2004) study the prediction properties of LASSO type estimators. For a high-dimensional setting, where the number of parameters can grow at a polynomial rate in the sample size $n$ and the true parameter vector is sparse in an $\ell_1$-sense, they show that

$$\mathbb{E}[(Y - \mathbf{x}^T \hat{\beta}_n)^2] - \sigma^2 \overset{\mathbb{P}}{\longrightarrow} 0 \quad (n \to \infty)$$

for a suitable choice of the penalty parameter $\theta = \theta_n$ (and other mild conditions), where $\sigma^2$ is the error variance. There are no strict conditions on the design matrix $X$. This risk consistency property is also called "persistency".

### 4.1.2. The Variable Selection Problem

An important theoretical question is: "Is it possible to determine the true model, at least asymptotically?" The answer is "Yes, but with some limitations". Meinshausen and Bühlmann (2006) show that LASSO is consistent for variable selection if and only if a neighborhood stability condition is fulfilled. Zhao and Yu (2006) made this condition more explicit and used the term "irrepresentable condition" for it. Under other assumptions, both sources show that LASSO is consistent for model selection, even if $p = p_n$ is allowed to grow (at a certain rate) as $n \to \infty$. The irrepresentable condition requires that the correlation between relevant and irrelevant predictors not be too large (we call a predictor relevant if the corresponding (true) coefficient is nonzero and irrelevant otherwise). Unfortunately, the theory assumes that the regularization parameter $\theta$ follows a certain rate, which is impractical for applications. Even so, the result implies that the true model is somewhere in the solution path with high probability. In practice, people often choose $\theta$ to be prediction optimal (or use some other criteria like $C_p$).

Meinshausen and Bühlmann (2006) and Leng *et al.* (2006) illustrate some situations where a prediction optimal selection of $\theta$ leads to estimated models that contain not only the true (relevant) predictors but also some noise (irrelevant) variables. For example, consider a high-dimensional situation with an underlying sparse model, that is where most variables are irrelevant. In this case a large value of the regularization penalty parameter $\theta$ would be required to identify the true model. The corresponding coefficients are biased significantly toward zero, and the estimator will have bad prediction performance. In contrast, a prediction optimal $\theta$ is smaller; in the resulting model, the relevant coefficients will not be shrunken too much, while the noise variables still have small coefficients and hence do not have a large effect on prediction.

Recently it has been shown that LASSO is consistent in an $\ell_q$-sense, for $q \in \{1, 2\}$. This means that

$$\|\hat{\boldsymbol{\beta}}_n - \boldsymbol{\beta}\|_q \xrightarrow{\mathbb{P}} 0 \quad (n \to \infty), \tag{16}$$

(Meinshausen and Yu 2008; Zhang and Huang 2007; Bunea *et al.* 2007; van de Geer 2008); for a high-dimensional setting and a suitable sequence $\theta_n$, often under much fewer restrictions than needed for model selection consistency. For fixed dimension $p$, this convergence result implies that coefficients corresponding to the relevant predictors will be non-zero with high probability. The conclusion is that the sequence of models found using LASSO contains the true model with high probability, along with some noise variables.

This suggests that LASSO be used as a 'variable filtering' method. When there are a very large number of predictors, a single regularization parameter $\theta$ is not sufficient for selecting variables and coefficient estimation. LASSO may be used to select a small set of predictors, followed by a second step (LASSO or otherwise) to select coefficients for those predictors, and also to perform additional variable selection in some cases.

### *4.2. Adaptive LASSO and related methods*

One example of a two-step method is *relaxed LASSO* (Meinshausen, 2007)[7]. It works roughly as follows: Calculate the whole path of LASSO solutions and identify the different submodels along the path. For each submodel, use LASSO again, but with a smaller (or no) penalty parameter $\phi\theta$, where $\phi \in [0, 1]$, i.e. no model selection takes place in the second step. By definition, relaxed LASSO finds the same sets of submodels as LASSO, but estimates the coefficients using less shrinkage: Model selection and shrinkage estimation are now controlled by two different parameters.

The hope is that the true model is somewhere in the first LASSO solution path. Relaxing the penalty may give better parameter estimates, with less bias toward zero. If we use $\phi = 0$ in the second step, this is exactly the LARS/OLS hybrid in Efron *et al.* (2004). In most cases, the estimator can be constructed at little additional cost by extrapolating the corresponding LASSO paths. Empirical and some theoretical results show the superiority over the ordinary LASSO in many situations. Meinshausen (2007) shows that the convergence rate of $\mathbb{E}[(Y - \mathbf{x}^T \hat{\boldsymbol{\beta}}_n)^2] - \sigma^2$ is mostly unaffected by the number of predictors (in contrast to the ordinary LASSO) if the tuning parameters $\theta$ and $\phi$ are chosen by cross-validation. Moreover, the conjecture is that a prediction-optimal choice of the tuning parameters leads to consistent model selection.

Another two-step method is *adaptive LASSO* (Zou, 2006). It needs an initial estimator $\hat{\beta}_{init}$, e.g. the least-squares estimator in a classical ($p < n$) situation.

---

[7]R package `relaxo` is available.

Weights can then be constructed based on the importance of the different predictors. For example, if the coefficient of the initial estimator is rather large, this would seem to indicate that the corresponding variable is quite important, and the corresponding coefficient shouldn't be penalized much. Conversely, an unimportant variable should be penalized more. The second step is a reweighted LASSO fit, using a penalty of the form

$$\theta \sum_{j=1}^{p} \hat{w}_j |\beta_j|,$$

where $\hat{w}_j = 1/|\hat{\beta}_{init,j}|^\gamma$ for some $\gamma > 0$. Note that the weights are constructed in a 'data adaptive' way. As with relaxed LASSO, the idea is to reduce bias by applying less shrinkage to the important predictors. From a theoretical point of view, this leads to consistent model selection, under fewer restrictions than for LASSO. If $\theta = \theta_n$ is chosen at an appropriate rate it can be shown that

$$\lim_{n \to \infty} \mathbb{P}[\hat{\mathcal{A}}_n = \mathcal{A}] = 1,$$

where $\hat{\mathcal{A}}_n$ is the estimated model structure and $\mathcal{A}$ is the true underlying model structure. As in all penalty methods, the choice of the penalty parameter $\theta$ is an issue, but prediction-optimal tuning parameter selection gives good empirical results.

Besides model selection properties, adaptive LASSO enjoys 'oracle properties': it is asymptotically as efficient as least squares regression using the perfect model (all relevant predictors and no others) as identified by an oracle:

$$\sqrt{n}(\hat{\boldsymbol{\beta}}_{n,\mathcal{A}} - \boldsymbol{\beta}_{\mathcal{A}}) \to \mathcal{N}(0, \sigma^2(\mathbf{C}_{\mathcal{A}\mathcal{A}})^{-1}) \quad (n \to \infty),$$

where $\mathbf{C}_{\mathcal{A}\mathcal{A}}$ is the submatrix of $\mathbf{C} = \lim_{n \to \infty} \frac{1}{n} \mathbf{X}^T \mathbf{X}$ corresponding to the active set.

Implementation of the adaptive LASSO estimator is easy: After a rescaling of the columns of the design matrix with the corresponding weights, the problem reduces to an ordinary LASSO problem. Huang *et al.* (2008) develop some theory about the adaptive LASSO in a high-dimensional setting. Several authors discuss applying the adaptive idea to other LASSO models and prove their oracle properties: Wang and Leng (2006) for group LASSO, Wang *et al.* (2007b) for autoregressive error models, Ghosh (2007) for elastic net, and Zhang and Lu (2007) and Lu and Zhang (2007) for Cox's proportional hazards model.

A predecessor of the adaptive LASSO is the nonnegative garrote (Breiman, 1995). It rescales an initial estimator by minimizing

$$\|Y - \sum_{j=1}^{p} \mathbf{x}_j \hat{\beta}_{init,j} c_j\|_2^2 + \theta \sum_{j=1}^{p} c_j,$$

subject to $c_j \geq 0$ for all $j$. Indeed, the adaptive LASSO with $\gamma = 1$ and the nonnegative garrote are almost identical, up to some sign constraints (Zou, 2006). The nonnegative garrote is for example also studied in Gao (1998) and Bühlmann and Yu (2006). More recently, Yuan and Lin (2007b) proved some consistency results and showed that the solution path is piecewise linear and hence can be computed efficiently.

The above methods try to reduce the bias of the estimates for the relevant predictors by applying less shrinkage whenever the corresponding coefficients are large. This raises the question of whether we could achieve similar behavior with a suitably chosen penalty function. Fan and Li (2001) propose the *SCAD* (smoothly clipped absolute deviation) penalty, a non-convex penalty that penalizes large values less heavily. It also enjoys oracle properties. The main drawback is the computational difficulty of the corresponding non-convex optimization problem.

Zou and Li (2008) make a connection between (adaptive) LASSO and SCAD. They use an iterative algorithm based on a linear approximation of the SCAD penalty function (or other penalties). In an approximation step, an (adaptive) LASSO problem is solved, and hence a sparse solution is obtained. This solution is then used for the next approximation step, and so on. However, it is not necessary to use more than one iteration: their *One-Step* (one iteration) estimator is asymptotically as efficient as the final solution, and hence also enjoys oracle properties.

Conversely, adaptive LASSO can also be iterated: the coefficients can be used to build new weights $\hat{w}_j$, and new coefficients can be calculated using these weights, and the iteration can be repeated. Bühlmann and Meier (2008) and Candes *et al.* (2007) find that doing multiple steps can improve estimation error and sparsity.

### *4.3. Dantzig selector*

Candes and Tao (2007) propose an alternative variable selection method called *Dantzig selector*, by optimizing

$$\min_{\beta} \|\mathbf{X}^T(\mathbf{Y} - \mathbf{X}\boldsymbol{\beta})\|_{\infty} \quad \text{subject to} \quad \|\boldsymbol{\beta}\|_1 \leq t.$$

They discuss an effective bound on the mean squared error of $\beta$, and the result can be understood as a deterministic version of (16). This procedure, which can be implemented via linear programming, may be valuable in high dimensional settings. In contrast, Tibshirani (1996) originally proposed a quadratic programming solution for LASSO, though the LAR implementation is more efficient.

However, Efron *et al.* (2007) and Meinshausen *et al.* (2007) argue that LASSO is preferable to the Danzig selector for two reasons: implementation and performance. Although Dantzig selector has a piecewise linear solution path (Rosset and Zhu, 2007), it contains jumps and many more steps, making it difficult to design an efficient path-following algorithm like the LARS implementation of LASSO. Furthermore, in their numerical results, they show that LASSO performs as well as or better than Dantzig selector in terms of prediction accuracy and model selection.

## 5. Software

There are a number of `S-PLUS` and `R` packages related to LARS, including: `brdgrun` (Fu, 2000), `elasticnet` (Zou and Hastie, 2005a), `glars` (Insightful Corportation, 2006), `glasso` (Kim *et al.*, 2005a), `glmpath` (Park and Hastie, 2006a), `grplasso` (Meier *et al.*, 2008), `lars` (Efron and Hastie, 2003), `lasso2` (Lokhorst *et al.*, 1999), `relaxo` (Meinshausen, 2007).

## 6. Conclusions and Future Work

LARS has considerable promise, offering speed, interpretability, relatively stable predictions, nearly unbiased inferences, and a nice graphical presentation of co-efficient paths. But considerable work is required in order to realize this promise in practice. A number of different approaches have been suggested, both for linear and nonlinear models; further study is needed to determine their advantages and drawbacks. Also various implementations of some of the approaches have been proposed that differ in speed, numerical stability, and accuracy; these also need further assessment.

Alternate penalties such as the elastic net and fused LASSO have advantages for certain kinds of data (e.g. microarrays and proteomics). The original LARS methodology is limited to continuous or binary covariates; grouped LASSO and LAR offer an extension to factor variables or other variables with multiple degrees of freedom such as polynomial and spline fits. Work is needed to further investigate the properties of these methods, and to extend them to nonlinear models.

Further work is also needed to address some practical considerations, including order restrictions (e.g. main effects should be included in a model before interactions, or linear terms before quadratic), forcing certain terms into the model, allowing unpenalized terms, or applying different levels of penalties to different predictors based on an analyst's knowledge. For example, when estimating a treatment effect, the treatment term should be forced into the model and estimated without penalty, while covariates should be optional and penalized.

Additional work is needed on choosing tuning parameters such as the magnitude of the $\ell_1$ penalty parameter in LASSO and other methods, the number of steps for LAR, and the multiple tuning parameters for elastic net and fused LASSO. Closely related is the question of statistical inference: is a larger model significantly better than a simpler model? Work is needed to investigate and compare model-selection methods including $C_p$, AIC, BIC, cross-validation, and empirical Bayes.

Work is also needed to develop estimates of bias, standard error, and confidence intervals, for predictions, coefficients, and linear combinations of coefficients. Are predictions sufficiently close to normally-distributed to allow for the use of $t$ confidence intervals? Does it even make sense to compute standard errors? Coefficients are definitely not normally distributed, due to a point mass at zero; but when coefficients are sufficiently large, might $t$ intervals still be useful, and how would one compute the standard errors?

The signal-to-noise ratio needs to be examined for the proposed methods, and alternatives compared. Evidence for a good signal-to-noise ratio would provide a strong impetus for their adoption by the statistical community.

Speed is also an issue, particularly for nonlinear models, and especially when cross validation is used for model selection or bootstrapping is used for inferences. In the linear regression case the cross-product matrices or $QR$ decomposition required for computations can be calculated in a single pass through the data. In contrast, for nonlinear models, fitting each subset of predictors requires multiple passes. Development of fast methods for nonlinear models is highly desirable.

Finally, to truly realize the promise of these methods beyond the domain of academic research, work is needed on usability issues. Implementations must be robust, numerical and graphical diagnostics to interpret regression model output must be developed, and interfaces must be targeted to a broad base of users.

We close on a positive note, with comments in the literature about LARS: Knight (2004) is impressed by the robustness of LASSO to small changes in its tuning parameter, relative to more classical stepwise subset selection methods, and notes "What seems to make the LASSO special is (i) its ability to produce exact 0 estimates and (ii) the 'fact' that its bias seems to be more controllable than it is for other methods (e.g., ridge regression, which naturally overshrinks large effects) ..." Loubes and Massart (2004) indicate "It seems to us that it solves practical questions of crucial interest and raises very interesting theoretical questions ...". Segal *et al.* (2003) write "The development of least angle regression (LARS) (Efron *et al.*, 2004) which can readily be specialized to provide all LASSO solutions in a highly efficient fashion, represents a major breakthrough. LARS is a less greedy version of standard forward selection schemes. The simple yet elegant manner in which LARS can be adapted to yield LASSO estimates as well as detailed description of properties of procedures, degrees of freedom, and attendant algorithms are provided by Efron *et al.* (2004)."

The procedure has enormous potential, which is evident in the amount of effort devoted to the area by such a large number of authors in the short time since publication of the seminal paper. We hope that this article provides a sense of that value.

Additional information, including software, may be found at www.insightful.com/lars